\documentclass [12pt] {article}

\newcommand {\adot} {\ensuremath{\dot a}}
\newcommand {\betai} {\ensuremath{\beta_i}}
\newcommand {\ehnukT}    {e^{h\nu/kT}}
\newcommand {\gbpbh}{\ensuremath{g_b^{BH}}}
\newcommand {\gbuni}{\ensuremath{g_b^{uni}}}
\newcommand {\gfpbh}{\ensuremath{g_f^{BH}}}
\newcommand {\gfuni}{\ensuremath{g_f^{uni}}}
\newcommand {\gspbh}{\ensuremath{g_\star^{BH}}}
\newcommand {\gsuni}{\ensuremath{g_\star^{uni}}}
\newcommand {\intnumin}{\ensuremath{\int_{\nu_{min}}^\infty}}
\newcommand {\intxmin} {\ensuremath{\int_{x_{min}}^\infty}}
\newcommand {\inu}  {\ensuremath{I_\nu}}
\newcommand {\mbh}  {\ensuremath{m}}
\newcommand {\mdot} {\ensuremath{\dot m}}

\newcommand {\mi}   {\ensuremath{m_i}}
\newcommand {\midot}{\ensuremath{\dot m_i}}
\newcommand {\mpbh} {\ensuremath{m}}
\newcommand {\mpl}  {\ensuremath{m_{pl}}}
\newcommand {\mprime}{\ensuremath{m^\prime}}
\newcommand {\mrel} {\ensuremath{m_{rel}}}
\newcommand {\myref}[1]	{(\ref{#1})}
\newcommand {\numin}{\ensuremath{\nu_{min}}}
\newcommand {\pR}   {\ensuremath{p_R}}
\newcommand {\rc}   {\ensuremath{r_c}}
\newcommand {\rhobh}{\ensuremath{\rho_{BH}}}
\newcommand {\rhoR} {\ensuremath{\rho_R}}
\newcommand {\rs}   {\ensuremath{R_S}}
\newcommand {\Tbh}  {\ensuremath{T_{BH}}}
\newcommand {\Tpbh} {\ensuremath{T_{BH}}}

\newcommand {\tpbh} {\ensuremath{t_{pbh}}}
\newcommand {\x}    {\ensuremath{\times}}
\newcommand {\xmin} {\ensuremath{x_{min}}}

\title {
   Baryogenesis from Primordial Blackholes after Electroweak Phase
   Transition
   }

\author {
   Niraj Upadhyay\thanks {E-mail: niraj@ducos.ernet.in},
   Patrick Das Gupta\thanks {E-mail: patrick@ducos.ernet.in}
   and
   R.~P.~Saxena\thanks {E-mail: rps@ducos.ernet.in}\\
   {\em Department of Physics and Astrophysics,}\\
   {\em University of Delhi,}\\
   {\em Delhi --- 110 007}\\
   {\em INDIA}
   }

\begin {document}

\maketitle

\begin{abstract}
%
%
Incorporating a realistic model for accretion of ultra-relativistic particles
by primordial blackholes (PBHs), we study the evolution of an
Einstein-de Sitter universe consisting of PBHs embedded in a thermal bath 
from the epoch $\sim 10^{-33}$ sec to $\sim 5\times 10^{-9}$ sec. In this 
paper we use Barrow et al's ansatz to model blackhole evaporation in which
the modified Hawking temperature goes to zero in the limit of the blackhole
attaining a relic state with mass $\sim m_{pl}$.
Both single mass PBH case as well as the case in which blackhole masses
are distributed in the range $8\times 10^2$ - $3\times 10^5$ gm have been
considered in our analysis. Blackholes with mass larger than $\sim 10^5$ gm
appear to survive beyond the electroweak phase transition and, therefore,
successfully manage to create baryon excess via $X-\bar X$ emissions, averting
the baryon number wash-out due to sphalerons. In this scenario, we find that 
the contribution to the baryon-to-entropy ratio by PBHs  of initial mass
$m$ is given by $\sim \epsilon \zeta (m/1\ {\mbox gm})^{-1}$, where $\epsilon$
and $\zeta$ are the CP-violating parameter and the initial mass fraction of
the PBHs, respectively.
For $\epsilon $ larger than $\sim 10^{-4}$, the observed matter-antimatter
asymmetry in the universe can  be attributed to the evaporation of PBHs.
   \end{abstract}

\begin{section} {Introduction}
%
%
That the Milky Way is essentially made of matter is evident not only
from the landings of space probes on Moon and other planets without
any disastrous consequences but also from the absence of anti-nuclei
in the observed cosmic rays, and from the observations of Faraday
rotation\cite{review:Steigmann}.
Observational support for absence of significant quantity of
anti-matter beyond our Galaxy exists, but it is of indirect nature
\cite{review:Steigmann}\cite{pap:DeRujula}.
Since visible mass in the universe is chiefly in the form of baryonic
matter, the inferred matter-antimatter asymmetry essentially boils
down to the problem of the origin of baryon asymmetry\footnote
{
   If neutrinos are massive then the gravitating mass may as well be
   dominated by leptons.
   However, there is hardly any direct measure of the lepton number of
   the universe.
}.
The baryon asymmetry is characterized by the baryon-to-photon ratio
$\eta = n_B/n_\gamma$, with $n_B$ and $n_\gamma$ being the number
densities of net baryons and photons, respectively.
According to standard big-bang nucleosynthesis calculations, the
predicted abundances of light elements depend only on the free
parameter $\eta$ and are in apparent agreement with the observed
abundances provided $\eta$ lies in the range $(2.8-4.5)\x 10^{-10}$
\cite{pap:OliveBBN}.
Recently, Tytler et al\cite{pap:TytlerEtAl} have estimated the
baryon-to-photon ratio from the observations of deuterium abundance in
a high red-shift quasar absorption system and according to their
measurements, log $\eta = -9.18 \pm 0.4 \pm 0.4 \pm
0.2$.

\par
The esthetically appealing scenario of the universe consisting of
equal amount of baryons and anti-baryons at the instant of creation is
still compatible with a non-zero $\eta$ if one invokes Sakharov
conditions, namely, of having $B$, $C$ and $CP$ violating interactions
in out-of-thermodynamic equilibrium condition sometime in the early
history of the universe \cite{pap:Sakharov}.
The Grand Unified Theories (GUTs) of fundamental forces incorporate
baryon number violating interactions naturally while $CP$ violation
can be introduced in such theories in many different ways (it is to be
noted that $CP$-violation added theoretically in GUTs, in general,
is not related to the observed CP-violation in the $K^\circ-\bar
K^\circ$ system\cite{GUTCP})
and therefore it is not surprising that GUTs provide a natural
framework for the generation of baryon asymmetry through decay of
$X-\bar{X}$ bosons\cite{paps:GUTBAU}.
However, through the work of Kuzmin, Rubakov and
Shaposhnikov\cite{pap:KuzminRubakovShaposhnikov} it came to be
appreciated that baryon number violation can take place during
electro-weak phase transition (EWPT) and that such processes could
erase baryon-asymmetry generated prior to EWPT era.
\par
Use of $B$-violation in electro-weak theories to produce excess
baryons has also been made in the literature\cite{review:EWBG} but one
of the major obstacles in this scenario is the requirement of low
Higgs mass which is in direct conflict with the experimental lower
limit of $m_H>88$ GeV\cite{ALEPH:NinetyEight}.
It appears that in the minimal version of electro-weak theory,
generating baryon asymmetry may not be possible at
all\cite{pap:ArchanEtAl}
and especially with the discovery of the top-quark with a mass around
175 GeV\cite{PDB:NinetySix} there is hardly any region left in the
parameter space of the standard model to produce observed
baryon-to-photon ratio\cite{KajanticEtAl}.

\par
The other major scenario of generating baryon asymmetry is to invoke
Hawking evaporation of black-holes.
The early sketchy ideas of Hawking and Zeldovich took proper  shape
with the advent of GUTs, giving rise to a picture of black-holes of
small mass emitting $X$ and $\bar{X}$ bosons thermally which
subsequently decay and in the process violate $B$, $C$ and $CP$,
leading to a production of baryon excess\cite{pap:PBHBAU}.
At the fundamental level, this scenario has an attractive feature in
that it combines ideas of black-hole
thermodynamics\cite{pap:Bekenstein}\cite{pap:HawkingEvaporation} on
one hand and GUT on the other, to explain the observed
matter-antimatter asymmetry in the universe.
One of the important ingredients of this picture is the occurrence of
mini-black-holes having mass less than $10^{14}$ gm.
It is obvious that such black-holes cannot emerge as end products of
stellar evolution.
However, Zeldovich and Novikov\cite{pap:PBHForm} and
Hawking\cite{pap:HawkingPBH} argued that primordial black-holes (PBHs)
of  small mass can be generated from the space-time curvature, and 
subsequently, Carr\cite{pap:Carr} showed the possibility of creating
PBHs from density fluctuations in the early universe.
In the context of inflation, several authors have discussed mechanisms to
produce PBHs using the general idea that bubble wall collisions may
trap pockets of false vacuum region that subsequently collapse to form
black-holes\cite{pap:FVPBH}. In a recent work, Nagatani
\cite{pap:Nagatani} has proposed an interesting {\em blackhole-electroweak}
 mechanism of  baryogenesis that requires the presence of a blackhole
 to create a domain wall around  it, leading to genesis of 
 baryon excess  without the need of a
 first order electroweak phase transition. 

\par
Previous paragraphs of this section indicate that although GUTs can
naturally generate baryon asymmetry, any baryon excess generated prior
to electro-weak era is erased due to sphaleron transitions, while at
the same time,
creation of baryon asymmetry solely due to electro-weak processes is
fraught with uncertainties as well as the requirement of low Higgs
mass, contrary to the experimental situation.
Under the present circumstances, it is therefore  natural to explore
alternate means to explain matter-antimatter asymmetry.
Since the existence of PBHs in the early universe is rather generic,
one ought to carefully re-examine the mechanism of generating baryon
asymmetry through black-hole evaporation.
In such a scenario, the crucial point to investigate is whether PBHs
survive after the EWPT has taken place, so that the baryon asymmetry
created due to their subsequent Hawking evaporation survives, leaving
an imprint till the present epoch.

\par
The present paper is an attempt to critically examine the evolution of
the masses of a collection of PBHs created after the end of
inflation, taking into account both the accretion of background matter
by the black-holes as well as the mass loss due to Hawking emission.
The paper has been organized in the following manner.
In Section 2 we discuss processes responsible for the change in a
black-hole's mass, and thereafter, we develop a formalism to describe
accretion of relativistic matter by mini-black-holes.
The subject of black-hole mass spectrum and its evolution is tackled
next, in section 3, along with a discussion on the cosmological
evolution of a mixture of PBHs and relativistic matter.
Section 4 deals with the study of evolution equations numerically as
well as a
detailed analysis of the numerical solutions pertaining to the
survival of PBHs past the EWPT.
In section 5, we calculate baryon excess resulting from the decay of
$X-\bar{X}$ bosons emitted by the PBHs during their final stages of
Hawking evaporation, and then discuss the implications of these results
to the question of matter-antimatter asymmetry.
Finally, we end with a brief discussion of the above scenario in
section 6.
   \end{section}

\begin{section} {Evolution of the mass of a black-hole}
   \label{sec:massevol}
   \begin{subsection}{Mass loss due to evaporation}
      \label{sec:mdoteva}
%
%
Bekenstein's conjecture\cite{pap:Bekenstein} that the area of the
event horizon of a black-hole being a measure of its entropy was
vindicated by the classic work of Hawking in the early seventies who
showed that when quantum effects around a black-hole are included, the
black-hole emits particles with a thermal distribution corresponding to
a temperature $T_{BH}$ that is proportional to the surface gravity at
the event horizon\cite{pap:Bekenstein}\cite{pap:HawkingEvaporation},
and is given by the relation,
\begin{equation}
   T_{BH} = \frac {\mpl^2}{8\pi\mbh} \frac{c^2}{k}
\label{eq:THawk}
\end{equation}
where {\mbh} and {\mpl} are the mass of the black-hole and the
Planck mass, respectively.
According to eq.~\myref{eq:THawk}, PBHs created in the early
universe with a mass
$\approx 10^{14}$ gm would be decaying today in a burst of high energy
radiation, and there exists in the literature, upper bounds on the
abundance of
such PBHs from the observed level of cosmic $\gamma$-ray
flux.
As pointed out by Zeldovich and others, the expression in
eq.~\myref{eq:THawk} for the
Hawking temperature can only be an
approximation and is amenable to modifications at Planck scale because
of the effects of quantum gravity.
In fact, particle physicists have shown from various angles that
Hawking evaporation may cease when the black-hole reaches the Planck
mass scale leading to a massive relic.
In this context, an interesting toy model inspired by superstring
theories has been considered by Barrow et al\cite{pap:BarrowRelic} in
which the expression for the black-hole temperature has been modified
by including correction terms that contain powers of black-hole mass
in units of Planck mass.
Following Barrow et al's ansatz, one can therefore express the
black-hole temperature as
\begin{equation}
   T_{BH} = \frac{\mpl}{8\pi} \left[ \frac{\mpl}{\mpbh} -
                                      \kappa\left(\frac{\mpl}{\mpbh}
                                            \right)^n
                              \right]
	 \frac{c^2}{k}
\label{eq:ModifiedTHawk}
\end{equation}
where $\kappa$ is a non-negative constant.
For $n>2$ and $\kappa\approx O(1)$, it is clear that for holes of mass
$m\gg\mpl$, eq.~\myref{eq:THawk} is a limiting case of
eq.~\myref{eq:ModifiedTHawk}.
According to eq.~\myref{eq:ModifiedTHawk}, as the hole mass decreases
due to evaporation, initially there is a rise in the hole's
temperature but as $\mbh$ approaches {\mpl} the temperature starts
falling and becomes zero when the mass of the hole reaches the value
$\mrel=\kappa^{1/(n-1)}\mpl$.
Therefore, Barrow et al's ansatz implies stable black-hole relics of
mass $\mrel\approx\mpl$.
To estimate the rate of mass-loss from eq.~\myref{eq:ModifiedTHawk}, we
may work in the frame-work of radiative transfer, assuming that the
hole's event horizon acts like a perfect black-body surface.
In such a case, it is easy to show that the energy flux $F$ is related
to the energy density $\varepsilon$ at the surface of interest in the
following manner\cite{book:RybickiLightman}
\begin{equation}
   F = \frac{c}{4} \varepsilon
\label{eq:Flux}
\end{equation}
The effective energy-density of ultra-relativistic particles due to
Hawking evaporation in the vicinity of the event horizon is related to
the temperature of the black-hole by
\begin{equation}
   \varepsilon = \frac {\pi^2 \gspbh}{30}
                 \frac {k^4}{(\hbar c)^3}
                 T_{BH}^4
\label{eq:energydensity}
\end{equation}
where $\gspbh = \gbpbh + (7/8)\gfpbh$ is the effective number of degrees
of freedom at the temperature $T_{BH}$, and {\gbpbh} and {\gfpbh} are the
corresponding degrees of freedom for bosons and fermions, respectively.
Therefore, the rate of mass-loss from the event-horizon is
given by
\begin{eqnarray}
   \frac{d\mbh}{dt} &=& - \frac{1}{c^2} F \cdot 4\pi\rs^2 \\
		    &=& - \alpha_2 m^2
			     \left[\frac{\mpl}{\mbh} -
                                  \kappa\left(\frac{\mpl}{\mbh}\right)^n
                             \right]^4
\label{eq:mdoteva}
\end{eqnarray}
where $\alpha_2 = \gspbh c^2/(30720\pi\hbar)$ and $\kappa \approx 1$.
In arriving at eq.~\myref{eq:mdoteva}, we have made use of
eqs.~\myref{eq:ModifiedTHawk}-\myref{eq:energydensity} as well as the
standard result for the Schwarzschild radius $\rs = 2G\mbh/c^2$.
The calculations that led to eq.\myref{eq:mdoteva} were based on modeling
the black-hole event-horizon to be the surface of a black-body of
radius {\rs} at a thermodynamic temperature {\Tbh}.
It is, therefore, interesting to compare our result with that of Don
Page\cite{pap:DonPage} which is based on rigorous numerical
computations for black-holes of mass $\mbh > 10^{17}$ gm.
According to his calculations, the mass-loss rate for such holes is
\begin{equation}
   \frac{d\mbh}{dt} = -2.011\x 10^{-4} \frac{\hbar c^4}{G^2\mbh^2}
\label{eq:DonPage}
\end{equation}
If we assume that eq.\myref{eq:DonPage} is valid also for $10^{-2}
\mbox{ gm} < m < 10^{17} \mbox{gm}$, then comparing \myref{eq:mdoteva}
and \myref{eq:DonPage} 
one obtains $\gspbh\approx 20$, which is not too unreasonable since
for holes of mass $10^{-2}$gm one expects {\gspbh} to be as high as
$\approx100$ (in most GUTs).
      \end{subsection}
   \begin{subsection}{Accretion of relativistic matter by a mini
                      black-hole}
%
%
\par
The temperature of the universe is expected to be extremely high just
after the end of inflation, and therefore matter during
that period will be in the form of ultra-relativistic particles.
For particles with de Broglie wavelength $\lambda \ll \rs$, 
the capture cross-section corresponding to a Schwarzschild blackhole
is $\sim\pi\rc^2$ where $\rc =
(3\sqrt{3}/2)\rs$\cite{book:LandauLifshitz}.
When the de Broglie wavelength of a particle is larger than $R_s$,
the capture cross-section is likely to be negligible as the blackhole
sees an incident wave rather than a point particle.
For high energy particles with $\lambda \ll \rs$, we will make use of
the geometric optics approximation in which any such
ultra-relativistic particle hitting a fictitious sphere of radius
{\rc} around the hole will be absorbed.

\par
If {\inu} represents the specific intensity of such particles
corresponding to energy $h\nu$ and if $dA$ is an area element on this
fictitious sphere then the rate at which energy is accreted by the
hole per unit range of $\nu$ per unit area is given by
\begin{eqnarray}
   \frac{dE_\nu}{dt d\nu dA} &= \int d\Omega \cos\theta \inu
                             &= \pi \inu
\end{eqnarray}
Since the effective area of capture is $4\pi\rc^2$, the rate at which
energy is accreted in the frequency range $[\nu,\nu+d\nu]$ is given by
\begin{equation}
   \frac{dE_\nu}{dt} = (2\pi\rc)^2 \inu d\nu
\label{eq:dEnudt}
\end{equation}

\par
To obtain the total rate of accretion of energy we integrate
eq.\myref{eq:dEnudt} over frequency keeping in mind that geometric optics
approximation requires the lower limit of integration $\nu_{min}$ to
be a few times $c/\rc$.
For ultra-relativistic particles, momentum is $p\approx h\nu/c$ so
that the number density of particles of species A in the
frequency-range $(\nu,\nu+d\nu)$ takes the form
\begin{equation}
   n_A(\nu)d\nu = \frac{4\pi g_A}{c^3}
		  \frac{\nu^2 d\nu}{\ehnukT \pm 1}
\label{eq:thermalnumden}
\end{equation}
where $g_A$ is the spin-degeneracy factor for the $A^{th}$ species and
the $+(-)$ sign refers to fermions (bosons).
In eq.~\myref{eq:thermalnumden} $T$ is the temperature of the universe.
Therefore, the specific intensity ${\inu}_A$ corresponding to the
species $A$ is given by\cite{book:RybickiLightman}
\begin{eqnarray}
   {\inu}_A &= \frac{ch\nu n_A(\nu)}{4\pi}
            &= \frac{g_A}{c^2} \frac{h\nu^3}{\ehnukT \pm 1}
\label{eq:thermalinu}
\end{eqnarray}
Making use of eqs.\myref{eq:dEnudt} and \myref{eq:thermalinu}, we can
express the net rate 
of energy accretion by a hole in the following manner
\begin{equation}
   \frac{dE}{dt} = \left(\frac{2\pi\rc}{c}\right)^2
		   \left[\gbuni\intnumin \frac{h\nu^3}{\ehnukT -1}
                                                   d\nu +
			 \gfuni\intnumin \frac{h\nu^3}{\ehnukT +1}
                                                   d\nu
                   \right]
\label{eq:dEdt}
\end{equation}
where $\numin=\alpha_1 c/\rc$ is the lower frequency cut-off,
$\alpha_1$ being a number of the order of 10 (this takes care of the
fact that only particles with $\lambda\ll\rs$ are considered to have 
been captured by the blackhole).
In eq.\myref{eq:dEdt}, \gbuni and \gfuni are the total bosonic and
fermionic degrees of 
freedom, respectively, for the cosmic soup.
These are to be distinguished from {\gbpbh} and {\gfpbh} introduced in
section \myref{sec:mdoteva}.

\par
The rate at which the hole's mass grows as a result of accretion is
\begin{equation}
   \frac{d\mbh}{dt} = \frac{405}{\pi^3 c^5} \varepsilon_R G^2 \mbh^2
      \left[\frac{\gbuni}{\gsuni}\intxmin \frac{x^3}{e^x -1} dx +
	    \frac{\gfuni}{\gsuni}\intxmin \frac{x^3}{e^x +1} dx
      \right]
\label{eq:mdotacc}
\end{equation}
where $\xmin = h\numin/kT$.

\par
In obtaining 
the above equation, we have made use of a change of variable in
eq.\myref{eq:dEdt} along with $\rc=(3\sqrt{3}/2)\rs$.
We note that $\varepsilon_R$ appearing in eq.\myref{eq:mdotacc} is the
energy-density of the background relativistic particles,
$\varepsilon_R = \pi^2\gsuni(kT)^4/(30\hbar^3c^3)$, {\gsuni} being the
temperature-dependent effective spin-degeneracy factor and is equal to
$\gbuni+7/8\gfuni$.
From eq.\myref{eq:mdotacc} it is evident that accretion plays an
important role for massive PBHs at early epochs when the temperature
of the universe is very high so that energy density $\varepsilon_R$ of
the relativistic particles is large while {\xmin} is small.
This is easy to understand from a physical point of view in the sense
that only when the temperature is large that there are sufficient
number of particles with de Broglie wavelength much less than the
Schwarzschild radius of the PBHs, ready to be accreted.
By the same token, when the hole-mass reaches a size of the order of
\mpl, neither accretion nor quantum evaporation is significant.
      \end{subsection}
\end{section}

\begin{section}{Blackhole mass spectrum and evolution of the universe}
   \label{sec:massspec}
%
%
It is evident that the mass distribution of PBHs is intimately linked
to the mechanism of their production.
Several authors
\cite{pap:Carr,pap:FVPBH,pap:NovikovEtAl,pap:Criticality}
in the literature have discussed blackhole mass spectrum from diverse 
angles.
Since the mass spectrum is sensitive to production mechanisms and,
since so far no particular model of PBH creation has been singled
out, we adopt a very general procedure in this paper to analyse the
evolution of blackhole mass spectrum.

\par
We consider a distribution function $N(m,t)$ such that $N(m,t)dm$
represents number-density of PBHs with mass in the range $(m,m+dm)$ at
the cosmic epoch $t$.
We assume that the creation of PBHs stopped after a cosmic epoch
{\tpbh} so that at later times in a given comoving volume the number
of holes remain the same while their masses change due to a
combination of Hawking radiation and accretion of background matter.
Note that we are working under the assumption that the ultimate state
of a PBH along the course of its evolution is a stable relic of mass
$\approx\mpl$, i.e. a hole does not disappear completely as the
original Hawking radiation mechanism would demand.
Also, since the mass $\mbh$ of a hole changes with time, the mass
distribution function at time $t$ and at time $t+dt$ are related as
\begin{equation}
   a^3(t) N(m,t) dm = a^3(t+dt) N(\mprime,t+dt) d\mprime
\label{eq:numconserve}
\end{equation}
where $\mprime$ is related to $m$ through $\mprime = m + \dot m dt$
and $a(t)$ is the FRW scale-factor at cosmic epoch $t$.
Making a Taylor expansion of quantities at the RHS of
eq.\myref{eq:numconserve}, and using the relation
\begin{equation}
   d\mprime  = dm \left( 1 + \frac{\partial \mdot}
		        {\partial m}dt
                  \right)
\end{equation}
we obtain
\begin{equation}
   \frac{\partial N}{\partial t} + 3\frac{\adot}{a}N +
      \frac{\partial}{\partial m}\left(N\mdot\right) = 0
\label{eq:Nevolve}
\end{equation}
With the help of the mass distribution function $N(m,t)$, we can also
obtain an expression for the mass-density associated with PBHs as
\begin{equation}
   \rhobh(t) = \int_{\mrel}^\infty m N(m,t) dm
\label{eq:rhoBH}
\end{equation}

\par
It is useful to express the black-hole mass-distribution as
\begin{equation}
   N(m,t) = N_0(t) f(m,t)
\label{eq:Nmt}
\end{equation}
where $N_0 \propto a^{-3}(t)$ so that $\int_{\mrel}^\infty f(m,t) dm$
is independent of time.
With the help of eq.\myref{eq:Nmt} it can be easily shown that
eq.\myref{eq:Nevolve} reduces to
\begin{equation}
   \frac{\partial f}{\partial t} + \frac{\partial}{\partial m} (\mdot
f) = 0
\label{eq:vlasov}
\end{equation}
Essentially, $f(m,t) dm$ represents the number of black-holes with
mass in the interval $(m,m+dm)$ in a unit coordinate volume at the
cosmic epoch $t$, while the dilution of black-hole number density due
to the expansion of the universe is taken care of by the factor
$N_0(t) = A/a^3(t)$.
Differentiating eq.\myref{eq:rhoBH} with respect to $t$ and then
making use of eqs.\myref{eq:Nmt} and \myref{eq:vlasov} it can be shown
that
\begin{equation}
   \frac{d\rhobh}{dt} + \frac{3\adot}{a}\rhobh = N_0 \int_{\mrel}^\infty
      \mdot f(m,t) dm
\label {eq:SpectrumFriedmann}
\end{equation}
Since the total energy-momentum tensor is divergence free, we also
have the equation\cite{book:Weinberg}:
\begin{equation}
   c^2 \frac{d}{dt}\left[(\rhoR+\rhobh)a^3\right] + 3p_Ra^2\adot = 0
\label {eq:DivTmunuZero}
\end{equation}
where $\rho_R = \varepsilon_R/c^2 $ is the mass density of radiation.
Here we have assumed that the black-holes possess negligible peculiar
speeds so that their contribution to pressure is insignificant.
Using $\pR = c^2\rhoR/3$ and eq.\myref{eq:DivTmunuZero} in
eq.\myref{eq:SpectrumFriedmann} we obtain:
\begin{equation}
   \frac{d\rhoR}{dt} + 4\frac{\adot}{a}\rhoR = -N_0 \int_{\mrel}^\infty
	\dot m f(m,t) dm
\label {eq:SpectrumFriedmannTwo}
\end{equation}
Eq\myref{eq:SpectrumFriedmannTwo} just reflects, as is to be expected,
the fact that an effective black-hole mass loss (or gain) would imply
$\rhoR \propto a^{-4-\alpha}$ where $\alpha(t)$ is negative(positive)
because of black-holes acting as source (sink) of radiation.

\par
In any mechanism of PBH production, the actual masses of the
black-holes will be distributed in a discrete fashion, and therefore
without loss of generality the distribution function can be expressed
as
\begin{equation}
   f(m,t) = \sum_{i=1}^K \betai \delta(m-\mi(t))
\label {eq:fmtOne}
\end{equation}
where {\betai} are constant weights corresponding to {\mi}, and $K$ is
the number of distinct black-hole masses.
It can easily be ascertained that the distribution function in
eq\myref{eq:fmtOne} indeed is a solution of
eq\myref{eq:vlasov}, since
\begin{equation}
   \frac{\partial}{\partial t} \delta(m-\mi(t)) =
\frac{\midot}{m-\mi(t)} \delta(m-\mi(t))
\end{equation}
and
\begin{equation}
   \frac{\partial}{\partial m} [\mdot f(m,t)] = - \sum_{i=1}^K \betai
\frac{\midot}{m-\mi(t)} \delta(m-\mi(t))
\end{equation}
Consequently, we have:
\begin{equation}
   \frac{d\rhoR}{dt} + 4\rhoR\frac{\adot}{a} = - N_0(t) \sum_{i=1}^K
      \betai \midot
\label{eq:rhor}
\end{equation}
The manner in which the individual mass {\mi} of a PBH changes with
time depends on the combination of Hawking evaporation rate and the
accretion of background relativistic matter as discussed in section
\ref{sec:massevol}.
Therefore, making use of eq \myref{eq:mdoteva} and eq
\myref{eq:mdotacc} in the context of a PBH with mass \mi we get the
following result:
\begin{eqnarray}
   \frac{d\mi}{dt} & = \frac{405}{\pi^3 c^3} \rhoR G^2 \mi^2
	\left[ \frac{\gbuni}{\gsuni}\intxmin \frac{x^3 dx}{e^x-1} +
               \frac{\gfuni}{\gsuni}\intxmin \frac{x^3 dx}{e^x+1}
        \right]
   \\\nonumber
   & -
   \alpha_2 \mi^2 \left[\frac{\mpl}{\mi} - \kappa
      \left(\frac{\mpl}{\mi}\right)^n\right]^4 \label{eq:midot}
\end{eqnarray}
From eqs.(17),(18) and (23), the mass density $\rho_{BH}$ associated
 with the PBHs can be written as:
\begin{equation}
   \rhobh(t) = N_0(t) \sum_{i=1}^K \betai \mi(t)
\end{equation}
The evolution of the scale-factor $a(t)$ then follows from the flat
FRW Einstein equation:
\begin{equation}
   \left(\frac{\adot}{a}\right)^2 = \frac{8\pi G}{3} \left[\rhoR +
N_0(t)\sum_{i=1}^K \betai \mi(t)\right]
\label{eq:sfactor}
\end{equation}
In writing down the above equation, we have adopted the inflationary
paradigm according to which universe in the post-inflationary
phase is described essentially by a flat FRW model.
In this paper, the evolution of the universe is determined by three
coupled differential equations (\ref{eq:rhor}),(27) and
(\ref{eq:sfactor}) along with the 
fact that $N_0(t) \propto a^{-3}(t)$.
   \end{section}

\begin{section}{Numerical Evolution}
   \label{sec:numevol}
%
%
\par
In this section, we solve  $2+K$ coupled
non-linear, first order differential equations \myref{eq:rhor},
(27)
and \myref{eq:sfactor}, set up in the preceding
section, numerically using
Hemming's fourth-order, double precision predictor-corrector method.
To begin with, we fix $N_0(t)$ by demanding that $\betai m_i(t_0)
N_0(t_0)$ represents the initial fraction $\zeta_i$ of total mass
density $\rho(t_0)$ that lies in blackholes having initial mass
$m_i(t_0)$ so that, 
\begin{equation}
   \label{eq:zibetarelation}
   \beta_i \mi(t_0) N_0(t_0) = \zeta_i \rho(t_0)
   \end{equation}
As $N_0(t) \propto a^{-3}(t)$, we have, from eq\myref{eq:zibetarelation},
\begin{equation}
   \label{eq:Nnoughtdefined}
   N_0(t) = \frac{a^3(t_0)}{a^3(t)} \frac{\zeta_i \rho(t_0)}{\beta_i
m_i(t_0)}
   \end{equation}
Since $N_0(t)$ is independent of $i$, we obtain the following
relation between $\zeta_i$ and $\beta_i$,
\begin{equation}
   \label{eq:zetaidefined}
   \zeta_i \propto \beta_i m_i(t_0)
   \end{equation}
where the constant of proportionality in eq\myref{eq:zetaidefined} can
be determined from the following identity,
\begin{equation}
   \rho_R(t_0) = \rho(t_0) - \sum_{i=1}^K \zeta_i \rho(t_0)
   \end{equation}
leading to the following expression,
\begin{equation}
   \mbox{proportionality constant} = 
      \left(\sum_{i=1}^K \beta_i m_i(t_0)\right)^{-1}
      \left(1 - \frac{\rho_R(t_0)}{\rho(t_0)}\right)
   \end{equation}
Substituting eq\myref{eq:Nnoughtdefined} in eqs \myref{eq:rhor} and
\myref{eq:sfactor}, we obtain,
\begin{equation}
\label{eq:rhortwo}
   \frac{d\rhoR}{dt} + 4\rhoR\frac{\adot}{a} = -
      \frac{a^3(t_0)}{a^3(t)} \rho(t_0)\sum_{i=1}^K
      \zeta_i \frac{\midot}{m_i(t_0)}
\end{equation}
\begin{equation}
\label{eq:sfactortwo}
   \left(\frac{\adot}{a}\right)^2 = \frac{8\pi G}{3} \left[\rhoR +
      \frac{a^3(t_0)}{a^3(t)} \rho(t_0) \sum_{i=1}^K \zeta_i
         \frac{\mi(t)}{m_i(t_0)} \right]
,
\end{equation}
respectively.

\par
In our original formulation (see section 3), the blackhole
initial mass-spectrum was completely specified by the set of numbers
$\{\betai, m_i(t_0); i=1,\dots K\}$.
Equivalently, since {\betai} and $\zeta_i$ are related by equation
\myref{eq:zetaidefined}, we may as well specify the spectrum by the
set $\{\zeta_i, m_i(t_0); i=1,\dots K\}$.

\par
For the purpose of numerical evolution, it is convenient to cast
equations 
(27)
, \myref{eq:rhortwo} and
\myref{eq:sfactortwo} in terms of dimensionless quantities defined
below
\begin{eqnarray}
      \tau         &=& t \sqrt{G\rho_0}                    \\
      \alpha(\tau) &=& \frac{a(t)}{a_0}                 \\
      R(\tau)      &=& \frac{\rho_R(t)}{\rho_0} \alpha^4   \\
      M_i(\tau)    &=& \frac{m_i(t)}{m_i(t_0)}
\end{eqnarray}
where $\rho_0 \equiv \rho(t_0)$ and $a_0 \equiv a(t_0)$.
In terms of the above quantities, the system of differential equations
assumes the following form:
\begin{eqnarray}
   \alpha' &=& \frac{1}{\alpha}\sqrt{\frac{8\pi}{3}(R+
		{\alpha\sum_{i=1}^K \zeta_i M_i})} \\
   R'      &=& - \alpha \sum_{i=1}^K \zeta_i M_i' \\
   M_i'    &=& m_i(t_0) M_i^2 (G\rho_0)^{-1/2}\\\nonumber
           & &
                        \left(
                           \frac{405}{\pi^3c^3}G^2\rho_0R\alpha^{-4}J_i
                           -
                           \alpha_2
\left[\left(\frac{m_{pl}}{m_i(t_0)} \frac{1}{M_i}\right)
                                             - \kappa
                                          \left(\frac{m_{pl}}{m_i(t_0)} \frac{1}{M_i} \right)^n
                                    \right]^4
                        \right)
\end{eqnarray}
where prime denotes differentiation with respect to $\tau$ and for
convenience we have introduced
\[
        J_i \equiv J(x_{0_i},T)
            =
\frac{\gbuni}{\gsuni}\int_{x_{0_i}}^\infty\frac{x^3\,dx}{e^x-1} + 
\frac{\gfuni}{\gsuni}\int_{x_{0_i}}^\infty\frac{x^3\,dx}{e^x+1}
\]
with
$$
   x_{0_i} = \frac{h}{kT} \left[\frac{\alpha_1 c}{r_{ci}}\right]
$$
and
$$
   r_{ci} = \frac{3\sqrt{3} G m_i(t)}{c^2}
$$
We choose $t_0$ to be the cosmic-epoch when inflation ends $\approx
10^{-33}$ sec., and set $\rho_0 = 10^{56}$ GeV$^4$, which is the
density expected at GUT scale. In our numerical evolution program, the
actual values used for the following parameters are listed below:
\begin{eqnarray*}
   \alpha_1 &= 10 \\
   \kappa   &= 0.1  \\
   n        &= 3 \\
   \gbpbh   &= \gfpbh \\
            &= \gbuni \\
            &= \gfuni \\
            &= 50
\end{eqnarray*}
\par
First we consider the case when $K=1$, i.e. at the end of inflation
a fraction $\zeta $ of matter lies in blackholes, all with initial mass $m_0=
m_0(t_0)$.
We study different models by varying $\zeta $ in the
range $10^{-3}$ to $10^{-1}$ while $m_0$ runs through the range $10^3$ to
$5\times 10^5$ gm.
In fig (1) we plot $a(t)$ for a typical choice of
$\zeta$ and $m_0$.
The plots of $a(t)$ for a radiation-dominated (RD) FRW universe
($a\sim t^{1/2}$) and a matter-dominated (MD) FRW universe ($a\sim
t^{2/3}$) are also given in the same figure.
The initial behaviour of the system is that of a RD universe but soon
the evolution of the scale-factor $a$ becomes similar to that in an MD
universe, and subsequently, as the PBHs evaporate, the dynamics becomes RD
again. This is because initially energy density of relativistic matter gets depleted
owing to accretion by PBHs, resulting in its decrease {\em faster}
than the kinematic rate $a^{-4}$ (see fig (1))
so that the dominant contribution from `dust' like blackholes drives a
faster expansion rate.
We have also compared our results with approximate estimates
 obtained by assuming that
$a(t)\sim t^{2/3}$ from $t=t_0$(end of inflation) to $t=t_{EWPT}$
(epoch of EWPT) and that $a(t)\sim t^{1/2}$ afterwards.
For the range of parameters $10^3 < m_0 < 10^5$ (gm) and $0.001 <
\zeta < 0.1$, the estimates agree with our numerical results
to within an order of magnitude.

\par
In fig (2) we plot a typical mass $m$ as a function of time.
It is evident from the figure that growth of blackhole mass due to
accretion takes place only in the initial period when the temperature
and density of the universe is very high. This is anyway expected
since the de Broglie wavelength $\lambda $ of a typical particle just after the
end of inflation is $\sim 10^{-28}$ cm, while the $R_S$ for a blackhole
of mass as low as $\sim $ 100 gm is  $\sim 10^{-26}$ cm leading to a
substantial accretion because of $\lambda < R_S$ criteria.
At intermediate times the curve flattens out reflecting a balance
between accretion and Hawking evaporation. During this phase, the
dynamics is essentially MD  since radiation loses out
in the competition because of the expansion of the universe as well as
its attenuation due to accretion by PBHs.

\par  
Towards the end, Hawking evaporation begins to
dominate the evolution of PBH mass as the accretion automatically gets
switched off due to the decrease in temperature and density of
background radiation. 
In fig (3) we plot, for a typical choice of parameters
$\zeta =$ 0.01 and $m_0 = 2.5\times 10^{5}$, the ambient temperature of the universe $T$ as well
as
the Hawking temperature {\Tpbh} of the black-hole.
The straight line portion of the curve has slope equal to $-2/3$.
Thus $T$ falls, at intermediate times, as if the dynamics of the
universe was akin to that of a MD universe.
At later times, when evaporation becomes the dominant process in the
evolution of the holes, the universe at first starts cooling at a slower rate,
 but eventually reheats  due to rapid evaporation of the
blackholes (the reheat portion is not included in the figure).
The point at which EWPT occurs is marked by an arrow in the figure,  
 (i.e. $T$ is $\sim $ 100 GeV at this instant of time) corresponding
to a value of $\sim 2\times 10^{-13} $ sec. We note that the epoch of
EWPT is considerably lower than the standard value of $\sim 10^{-10}$
sec obtained from the time-temperature relation in big-bang
models. The reason for this is not hard to understand, as depletion of
radiation by the accreting PBHs leads to a MD phase causing the $T$ to
decline faster than the usual $t^{-1/2}$ fall.

Now, the  amount of reheating should be such that the temperature of the
universe does not rise above the EWPT temperature ($\sim $ 100 GeV),
 because, otherwise
the sphaleron processes will be re-ignited, leading once again to a
 washing out of  BAU generated.
This, in effect, constrains our parameters $\zeta$ and $m_0$.
In fig (4) we plot the combination of $\zeta$ and $m_0$
for which the re-heat temperature is 100 GeV, and these points are
empirically fitted with a curve. From the numerical evolution, we find
 that the region lying below to the right of the curve consists
 of those values of ($\zeta $,$m_0$) for which
 re-heat temperature remains below 100 GeV. While the region
lying left of the curve consists of those combinations for which PBHs
evaporate away, reaching the relic state before EWPT, and therefore
are not of any use as far as baryogenesis is concerned. From
 fig (4) it is evident that for $\zeta $ lying in the interval
   ($10^{-5}$,0.1) a PBH with initial mass less than $\sim 2 \times 10^5 $ gm
 converges to the relic state before EWPT, and hence does not contribute
 to  generation of baryons. Therefore, we find that if the initial
 PBH mass spectrum is a delta-function peaking at the mass $m_0 $,
 baryogenesis through blackhole evaporation is viable only when the initial
 mass of the PBHs exceeds $\sim 2\times 10^5 $ gm for reasonably low values of
 $\zeta $.  

\par
Next, we consider the case in which  blackhole masses at time
 $t_0$ are distributed in a pseudo-Maxwellian manner as shown in
 fig (5). The blackhole masses fall in a range from
$8\times 10^2$ gm to $3\times 10^5$ gm, with 22 distinct mass values 
contributing to a total fraction $\sum_i^{22} \zeta_i \sim $ 0.09 of the
mass density of the universe just after the end of inflation. From the
plot fig (6) it is apparent that blackholes with larger initial mass accrete
background hot matter at higher rates than those with smaller initial
mass, as expected from the fact that higher mass PBHs have larger cross-section
for absorbing matter.  We find that those PBHs with initial
 mass greater than $1.5\times 10^5$ gm
reach $X-\bar X$ emitting phase after the epoch $1.9\times 10^{-11}$ sec,
 the instant at which EWPT takes place for this spectrum of masses.
Once again we find that EWPT occurs sooner than that in the standard
model. There
 are 7 of such blackhole masses which finally contribute to the production 
of baryon excess. Because of the wide distribution of blackhole masses, the
instants at which the PBHs reach the relic mass are staggered,
 hence no sharp re-heating takes place in our analysis, rather
 the temperature of the universe falls at a slower rate till the largest
  size blackhole (with initial mass $= 3\times 10^5$ gm) evaporates, leaving
 behind a relic mass  around the epoch $\sim 3\times 10^{-9}$ sec,
when the temperature of the universe is $\sim $ 9 GeV.
The decline of  temperature with time is shown in
fig (7).

\par
Even in the case
of blackhole mass distribution, $K$ being
larger than 1, in principle, one can constrain the parameter space 
$(\zeta_i,m_i(t_0))$ from the requirement of re-heating less than 100
GeV (as undertaken when $K=1$, see fig (4)), however the
exercise is enormously time consuming, and is beyond the scope of the
 present paper.

   \end{section}

\begin{section}{Baryogenesis}
%
%
We saw in the previous section that for $\zeta\approx 0.01$,
blackholes created with mass less than $\approx 2\x 10^5$ gm evaporate
and reach the relic state before the EWPT and hence their contribution
to baryon asymmetry is doubtful due to the expected $B$-violation
induced by sphalerons.
However, blackholes with initial mass larger than $\approx 2.5\x 10^5$
gm certainly ought to be considered as sources of baryogenesis since
they reach the $X-\bar X$ emission phase well past the EWPT.
In this section, we proceed to estimate the quantity of excess baryons
resulting from blackholes whose Hawking temperature reaches GUT scale
after EWPT.

\par
Representing the specific intensity of $X$-bosons radiated with energy
$h\nu$ from a blackhole by $I^X_\nu$, we have the relation (e.g., see
\cite{book:RybickiLightman})
\begin {equation}
   I^X_\nu = \frac{u^X_\nu(\Omega)}{v}
\end {equation}
where $u^X_\nu(\Omega)$ is the specific energy density and $v$ is the
speed of the emanating $X$-bosons.
 
With
\begin {equation}
   v = c \left[ 1 - \left(\frac{mc^2}{h\nu}\right)^2 \right]^{1/2}
\end {equation}
and
\begin{equation}
   u^X_\nu(\Omega) = \frac{h\nu^3}{c^4} \frac{v g_X}{e^{h\nu/kT_{BH}(m)}-1}
\end{equation}
we may express $I^X_\nu$ as
\begin {equation}
   I^X_\nu = \frac{h\nu^3}{c^2} \left[ 1 -
      \left(\frac{mc^2}{h\nu}\right)^2 \right]
      \frac{g_X}{e^{h\nu/kT_{BH}(m)} -1}
\label {eq:a}
\end {equation}
It is to be noted that $g_X$ and $T_{BH}(m)$ are the spin degeneracy factor
of $X$-bosons and Hawking temperature of a blackhole of mass $m$,
respectively.

\par
The flux-density of $X$-bosons at a distance $r$ from the blackhole is
given by
\begin{equation}
   F^X_\nu = \pi I^X_\nu \frac{\rs^2}{r^2}
\label {eq:b}
\end{equation}
Therefore, from equations \myref{eq:a} and \myref{eq:b} the rate of
emission of $X$-bosons from a blackhole of mass m is derived to be
\begin{eqnarray}
   \frac{dN_X(m)}{dt} &=& \int_{m_Xc^2/h}^\infty
   \frac{F^X_\nu \cdot 4\pi r^2 d\nu}{h\nu}\\
    &=& \frac{4\pi\rs^2 c^4 g_X m^3_X}{h^3} I(y_i)
\label{eq:c}
\end{eqnarray}
where $m_X$ is the mass of the $X$-boson and
\begin{equation}
   I(y_i) = \int_1^\infty \frac{y^2-1}{e^{y/y_i}-1} dy
\label{eq:Iyi}
\end{equation}
while
\begin{equation}
   y_i(t) \equiv \frac{kT_{BH}(m_i(t))}{m_Xc^2}
\label{eq:yi}
\end{equation}
Since, at any given cosmic epoch $t$, the number density of blackholes
with mass lying in the interval $(m,m+dm)$ is $N_0(t) f(m,t)$ (see eq
\myref{eq:Nmt}), the rate at which $X$ and $\bar X$ bosons are
generated in a unit proper volume is given by
\begin{equation}
   \frac{dn_{X\bar X}}{dt} = 2 N_0(t) \int \frac{dN_X(m)}{dt} f(m,t)
dm
\label{eq:d}
\end{equation}
In eq \myref{eq:d} the factor $2$ arises because we have included
production of $\bar X$-bosons as well.
Making use of the form given in eq \myref{eq:fmtOne} we can express eq
\myref{eq:d} as
\begin{equation}
   \frac{dn_{X\bar X}}{dt} = 2 N_0(t) \sum_{i=1}^K \betai
      \frac{dN_X(m_i)}{dt}
\label{eq:e}
\end{equation}
The lifetime of a $X$-boson $\tau_X = \Gamma_X^{-1}$ turns out to be
$\approx 10^{-36}$ sec when $m_X\approx 10^{14}$ GeV
\cite{pap:KolbRiottoTkachev} which is negligible in comparison with
the time scales over which blackhole mass changes or the universe
expands appreciably.
Hence, the rate of increase of net baryon number in a unit proper
volume is
$$
   \approx \epsilon \frac{dn_{X\bar X}}{dt}
$$
with
\begin{equation}
   \epsilon \equiv \frac{\Gamma(X\to q l) - \Gamma(\bar X \to \bar q \bar
              l)}
              {\Gamma_{tot}}
\label{eq:f}
\end{equation}
being the net baryon number generated by the decay of a pair of $X$
and $\bar X$ \cite{book:KolbAndTurner}.

\par
If $n_B(t)$ represents  net baryon number density at the cosmic
epoch $t$ then
\begin{equation}
   \frac{d}{dt}(a^3(t)n_B(t)) = \epsilon a^3(t) \frac{dn_{X\bar
X}}{dt}
\label{eq:g}
\end{equation}
Employing eq \myref{eq:Nnoughtdefined} and \myref{eq:e} in
\myref{eq:g} and then integrating the latter, we obtain
\begin{equation}
   a^3(t)n_B(t) - a^3(t_{EWPT})n_B(t_{EWPT}) = 2 \epsilon (\rho_0
     a^3_0) \sum_{i=1}^K
   \frac{\zeta_i}{m_i(t_0)} \int_{t_{EWPT}}^t
\frac{dN_X(m_i)}{dt^\prime} dt^\prime
\label{eq:h}
\end{equation}
Assuming that prior to blackhole baryogenesis, the net baryon number
in the universe is zero (i.e. $n_B(t_{EWPT})=0$) and making use of eq
\myref{eq:c} in eq \myref{eq:h}, we get the following expression for
the net baryon number density at any time,
\begin{eqnarray}
   n_B(t) &= \frac{\rho_0 a_0^3}{a^3(t)} \left (\frac{4 G^2}{\pi \hbar^3}\right
)\cdot\epsilon g_X m^3_X\cdot
    \sum_{i=1}^K \frac{\zeta_i}{m_i(t_0)} \int_{t_{EWPT}}^t dt^\prime
m_i^2(t^\prime) I(y_i(t^\prime))
\label{eq:i}
\end{eqnarray}
After EWPT has taken place, the evolution of PBH mass is totally dominated
by eq \myref{eq:mdoteva} since the de Broglie wavelength $\lambda $ of
a typical particle is larger than $\sim 10^{-15}$ cm, while the $R_S$
corresponding to a blackhole  of mass as high as $\sim 10^7$ gm is
only $\sim 10^{-21}$ cm. Hence, using eq \myref{eq:mdoteva} we can
change the variable of integration in eq\myref{eq:i} from $t^\prime $ to
$m_i(t^\prime)$ so that,
\begin{equation}
   \int_{t_{EWPT}}^t m^2_i(t^\prime)I(y_i(t^\prime)) dt^\prime =
   \frac{2m_{pl}^4}{\alpha_2(8\pi m_X)^4}
   H({m_i(lower)},{m_i(t_{EWPT})})
\label{eq:j}
\end{equation}
where $H$ is defined to be,
\begin{equation}
   H({m_i(lower)},{m_i(t_{EWPT})})\equiv
   \int_{m_i(lower)}^{m_i(t_{EWPT})} \frac {1}{y_i^2}
   \left[
      y_i \sum_{k=1}^\infty \frac{e^{-k/y_i}}{k^3} +
          \sum_{k=1}^\infty \frac{e^{-k/y_i}}{k^2}
   \right] dm_i
\label{eq:k}
\end{equation}
In obtaining eqs \myref{eq:j} and \myref{eq:k} we have used the series
equivalent of the integral given in eq\myref{eq:Iyi}.The value of
 $m_i(lower)$ is set by requiring $y_i$ to be $10^{-3} $ since the
series given in eq\myref{eq:k} is negligibly small for smaller values
of $y_i$. This automatically takes into account the fact that only 
those PBHs matter for BAU that are capable of emitting $X-\bar X$
after EWPT.
For PBH
masses larger than $10^5 m_{pl}$, the value of $H$ is $2.7\times 10^{-2}$
and becomes insensitive to the exact value of $m_i(t_{EWPT})$
thereafter. Therefore, for the 7 PBHs that survive the EWPT, we
have,
\begin{equation}
   \sum_{i=16}^{22} \frac{\zeta_i}{m_i(t_0)} H(m_i(lower),
m_i(t_{EWPT}))
   = 1.97\times 10^{-9}
\label{eq:l}
\end{equation}

\par
The entropy density of the universe at any epoch $t$ is given by,
\begin{equation}
   s = \frac{2\pi^2}{45} \gsuni \frac{k^4}{(\hbar c)^3} T^3(t)
\label{eq:entropy}
\end{equation}
We estimate the baryon-to-entropy ratio at $t \sim 3\times 10^{-9}$
sec, when all the PBHs settle on to the relic state, by making use
of eqs\myref{eq:i},\myref{eq:j},\myref{eq:k},\myref{eq:l} and
\myref{eq:entropy}, 
\begin{equation}
   \frac{n_B(t)}{s(t)} = 7.5\times 10^{-8} \epsilon g_X
\left(\frac{\gspbh}{100}\right)^{-1}
\left( \frac{\gsuni}{100}\right)^{-1}
\label{eq:nbbys}
\end{equation}
The contribution to baryon-to-entropy ratio by PBHs with initial mass
$m_0$ and initial mass fraction $\zeta $  goes roughly as,
\begin{equation}
   \frac{n_B}{s} \approx \epsilon \zeta g_X
      \left(\frac{m_0}{1\ {\mbox gm}}\right)^{-1}
      \left(\frac{\gspbh}{100}\right)^{-1}
      \left( \frac{\gsuni}{100}\right)^{-1}
\label{eq:nbbys2}
\end{equation}
Hence, in the case of a delta-function mass spectrum with
 $\zeta \approx $ 0.01 and $m_0 \approx 2.5 \times 10^{5} $ gm,
one obtains a baryon-to-entropy ratio of $\approx$ $4\times 10^{-8}\epsilon$,
with $g_X=1$. Thus one may use eq\myref{eq:nbbys2} along with the
  value of $n_B/s \approx 10^{-11}$, that follows from observations, to
 put a constraint on $\epsilon\zeta/m_0$. This implies that one requires
 the CP-violating parameter $\epsilon $ to be around $\sim 10^{-4}$ to 
generate excess baryons from evaporating PBHs.
   \end{section}

\begin{section}{Discussions}
%
%
\par
To study the evolution of PBHs, in the early universe, that undergo
accretion along with steady mass loss due to Hawking evaporation,
 we have laid down a formalism which can handle any blackhole mass
spectrum that can be decomposed as a sum of weighted
$\delta$-functions. Accretion of ambient hot matter by a blackhole has
been 
modeled in the limit of geometric approximation, so that only those
particles with de Broglie wavelength less than about a tenth of
Schwarzschild 
radius are considered for absorption by the blackhole. The evolution
of a flat FRW universe and the PBHs has been studied numerically to
find conditions under which blackholes survive past the electroweak phase
transition in order that their subsequent evaporation leads to
baryogenesis.
\par       
The basic picture which emerges is the following. In the case of
a blackhole mass spectrum that peaks sharply at a single mass value $m_0$,
when $\zeta$ (the initial mass fraction of PBHs) is of the order of 1\%,
PBHs with initial mass $m_0$  less than about $2.3\times 10^5$
gm  evaporate {\em before} EWPT.
Therefore, only PBHs with $m_0$  greater than this critical value 
need be considered for generation of BAU.
Here, we wish to point out that the model of accretion which
 one considers can make an immense difference in the final
 result of the analysis.
 If one uses a simple spherical model of accretion in which the
capture-cross section is just $\pi R_S^2$ and with no de Broglie
wavelength based cutoff then blackholes of initial mass $m_0 \approx 10^3$
gm can successfully live past the EWPT, and eventually contribute to
the  BAU (see Majumdar et al. in \cite{pap:PBHBAU}).
While on using the same set of parameters with a wavelength based
cutoff model of accretion, we find that PBHs of such small initial mass
do not survive beyond the EWPT.

\par
For reasonable choice of parameters, we find that in the case of PBHs
with a distribution of mass ranging from $8\times 10^2$ - $3\times
10^5$ gm, blackholes with initial mass larger than about $\sim 10^5 $
gm reach the relic state much after EWPT. Because of the presence of
blackholes with mass less than $10^5$ gm that evaporate at a faster
rate, pumping in  energetic particles into the surrounding  medium, 
 the ambient temperature in this case declines at a slower rate, and
 hence EWPT takes place  later than in the case when all PBHs had the
 same mass of $2.5\times 10^5$ gm. As described in sections
\myref{sec:massspec} and \myref{sec:numevol}, the evolution of mass
spectrum is totally determined by the manner in which individual blackhole
masses change with time, $\beta_i$ or equivalently $\zeta_i$ remaining
fixed for all times. As an illustration, we have shown the evolution
of mass spectrum in fig (6) for a particular set of
$\zeta_i$.

\par
 We wish to point out that accretion is important only
during the initial stages just after the end of inflation when the
temperature of the universe is $\sim 10^{13} $ GeV, causing an increase
in the mass of a blackhole by a factor of $\sim 4$. There are two factors
responsible for a blackhole of initial mass of $\sim 10^5$ gm 
 to live after the EWPT. One being the increase in the mass due to
accretion, while the other is the 
occurance of EWPT sooner than that in a model in which there
 is no depletion of
radiation due to PBHs acting as sinks. For blackholes with mass less
than $\sim 10^5 $ gm, accretion is less due to the reduction in
capture cross-section because of which the rate of depletion of
radiation is not large leading to a delayed occurance of EWPT, after
the blackholes have reached the final relic state.
\par
Barrow et al's ansatz\cite{pap:BarrowRelic} which has been used in
this paper to take into account expected modification of Hawking
emission becomes important only when the mass of evaporating blackholes 
fall below $\sim 10 m_{pl}$. Our numerical results are not 
sensitive to the exact form of modified blackhole temperature. For
baryogenesis, significant quantity of $X-\bar X$ are emitted only during
the phase when blackhole temperature is $\sim T_{GUT}$, because of
which the integral $H$ is not sensitive to the upper limit
$m(t_{EWPT})$ so long as the latter is larger than $10^5\ m_{pl}$.
Therefore, the final expression for baryon-to-entropy ratio turns out
to be  rather simple (see eq \myref{eq:nbbys2}), implying that if at the 
end of inflation 1\%
of total matter goes into creating PBHs with
initial mass $2.5\times 10^5$ gm then this scenario can successfully
lead to BAU provided the CP-violating parameter $\epsilon $ is over 
$10^{-4}$. Thus production of baryon excess through blackhole
evaporation is a viable alternative to GUTs or electroweak baryogenesis,
although there is no denying that because of the presence of
parameters like $\zeta_i $ and $m_i(t_0)$ whose values {\em a
priori} are uncertain, this scenario cannot provide meaningful constraint
on the value of $\epsilon $.
   \end{section}

\begin{section}{Acknowledgements}
%
%
We wish to thank Dr.~Amitabha Mukherjee and Dr.~Archan Majumdar for
useful suggestions. It is a pleasure to thank Harvinder Kaur Jassal,
Hatem Widyan and Abha Dev for going through the {\em LATEX} file of
this paper carefully.
One of us (NU) would like to thank the University Grants Commission,
New Delhi, for financial support. 
   \end{section}

\bibliographystyle{plain}

%
\pagebreak
\pagestyle {empty}
\centerline {\large \bf Figure Captions}
\begin {enumerate}
   \item {Figure 1:}
      The evolution of the scale-factor $a(t)$ for
      $\zeta=0.01$, $m_0 = 2.5\x10^5$ gm.
      The plots $a\sim t^{1/2}$ and $a\sim t^{2/3}$ are provided for
      comparision.
   \item {Figure 2:}
      The evolution of the mass $m(t)$ of the PBHs for a
      typical choice $\zeta=0.01$, $m_0 = 2.5\x 10^5$ gm.
   \item {Figure 3:}
      The temperature $T$ of the background thermal bath
      and the Hawking temperature $T_{BH}$ for a typical choice $\zeta=0.01$
      and $m_0 = 2.5\x 10^5$ gm. The instant of EWPT is marked by an arrow.
   \item {Figure 4:}
      The combinations $\zeta$ and $m_0$ for which the reheat
      temperature $ = T_{EWPT} = 100$ GeV.
      The region with {\em acceptable} reheat temperatures $ <
      100$ GeV is indicated in the figure.
      The analytical fit with dotted line is {\em purely} empirical.
   \item {Figure 5:}
      Black-hole mass spectrum: plot of $\zeta_i$ against
      $m_i(t_0)$.
   \item {Figure 6:}
      The evolution of the masses $m_i(t)$ of a collection
      of PBH masses distributed according to the spectrum shown
      in  fig (5).
   \item {Figure 7:}
      The cooling of the universe for the case where PBH
      masses are distributed according to the spectrum
      displayed in fig (5).
      The epoch of EWPT is marked by an arrow, and it
      takes place at $1.9\times 10^{-11}$ sec.
   \end{enumerate} 
%
%
\newpage
\begin {figure}[ht]
 \vskip 15truecm
      \includegraphics{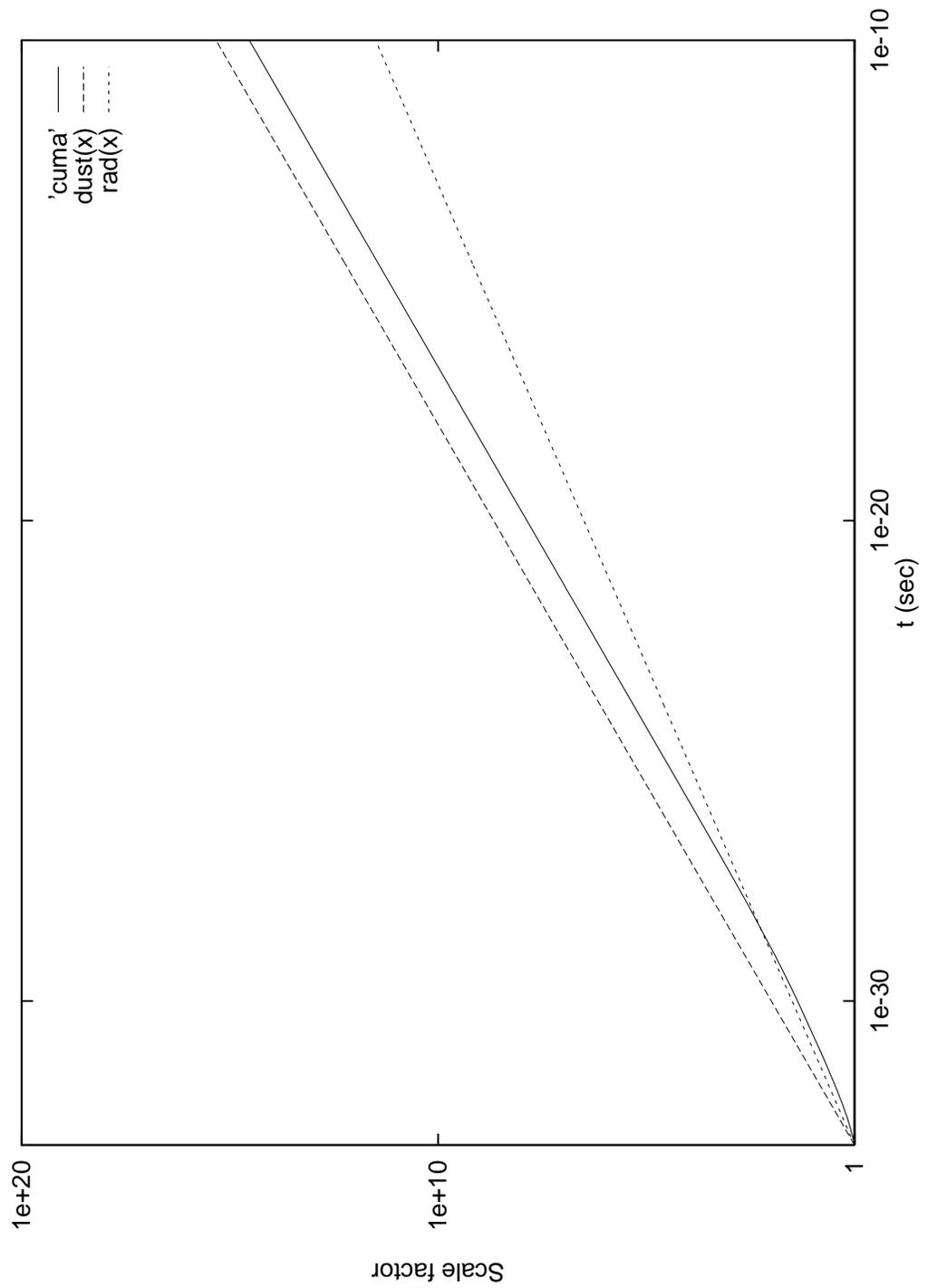}
   \caption {The evolution of the scale-factor $a(t)$ for
      $\zeta=0.01$, $m_0 = 2.5\x10^5$ gm.
      The plots $a\sim t^{1/2}$ and $a\sim t^{2/3}$ are provided for
      comparision.}
   \label {fig:ScaleFactor}
\end {figure}
\newpage
\begin {figure}[ht]
 \vskip 15truecm
      \includegraphics{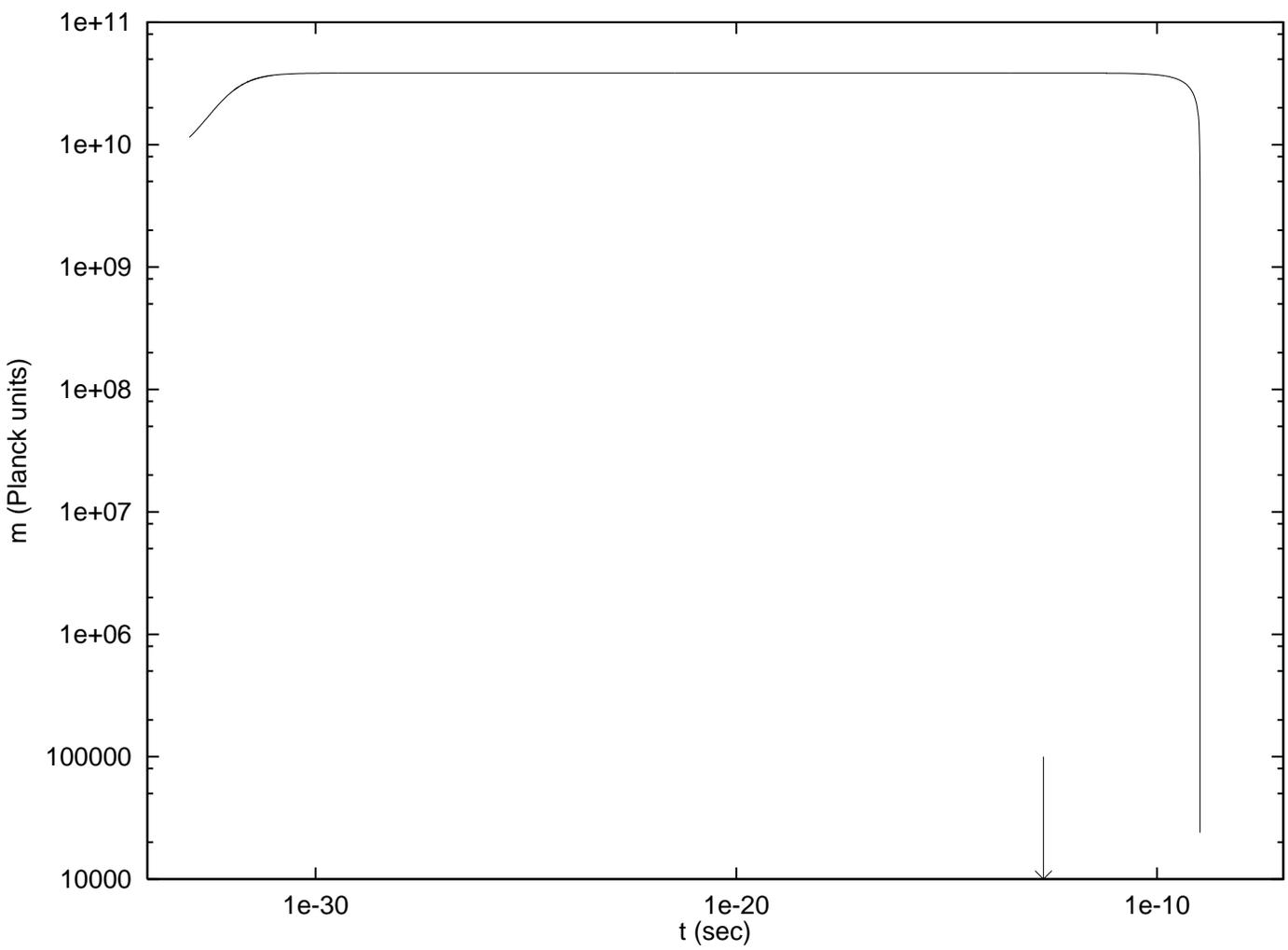}
   \caption {The evolution of the mass $m(t)$ of the PBHs for a
      typical choice $\zeta=0.01$, $m_0 = 2.5\x 10^5$ gm.}
   \label {fig:m}
\end {figure}
\newpage
\begin {figure}[ht]
 \vskip 15truecm
      \includegraphics{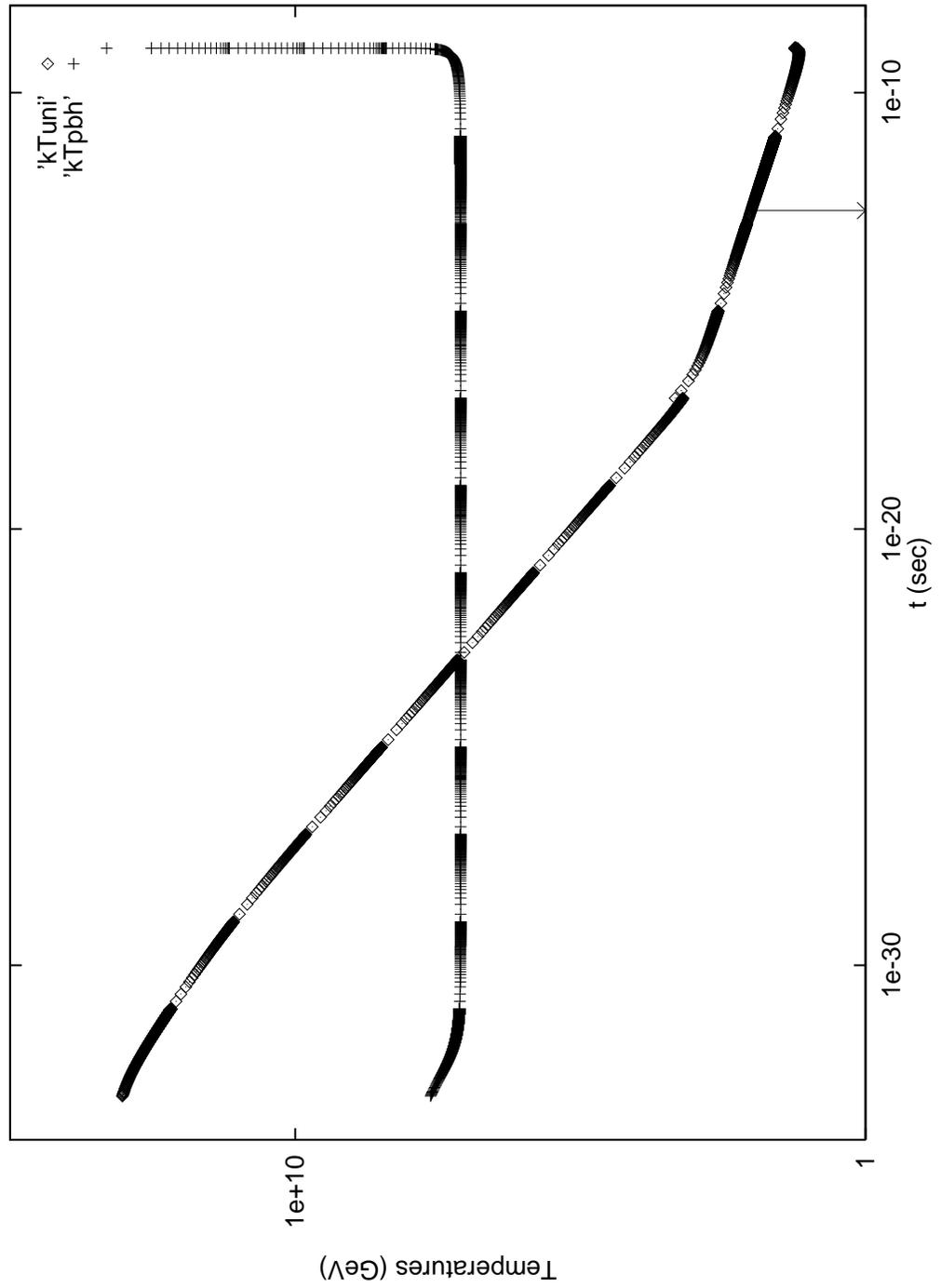}
   \caption {The temperature $T$ of the background thermal bath
      and the Hawking temperature $T_{BH}$ for a typical choice $\zeta=0.01$
      and $m_0 = 2.5\x 10^5$ gm. The instant of EWPT is marked by an arrow.}
   \label {fig:Temperatures}
\end {figure}
\newpage
\begin {figure}[ht]
 \vskip 15truecm
      \includegraphics{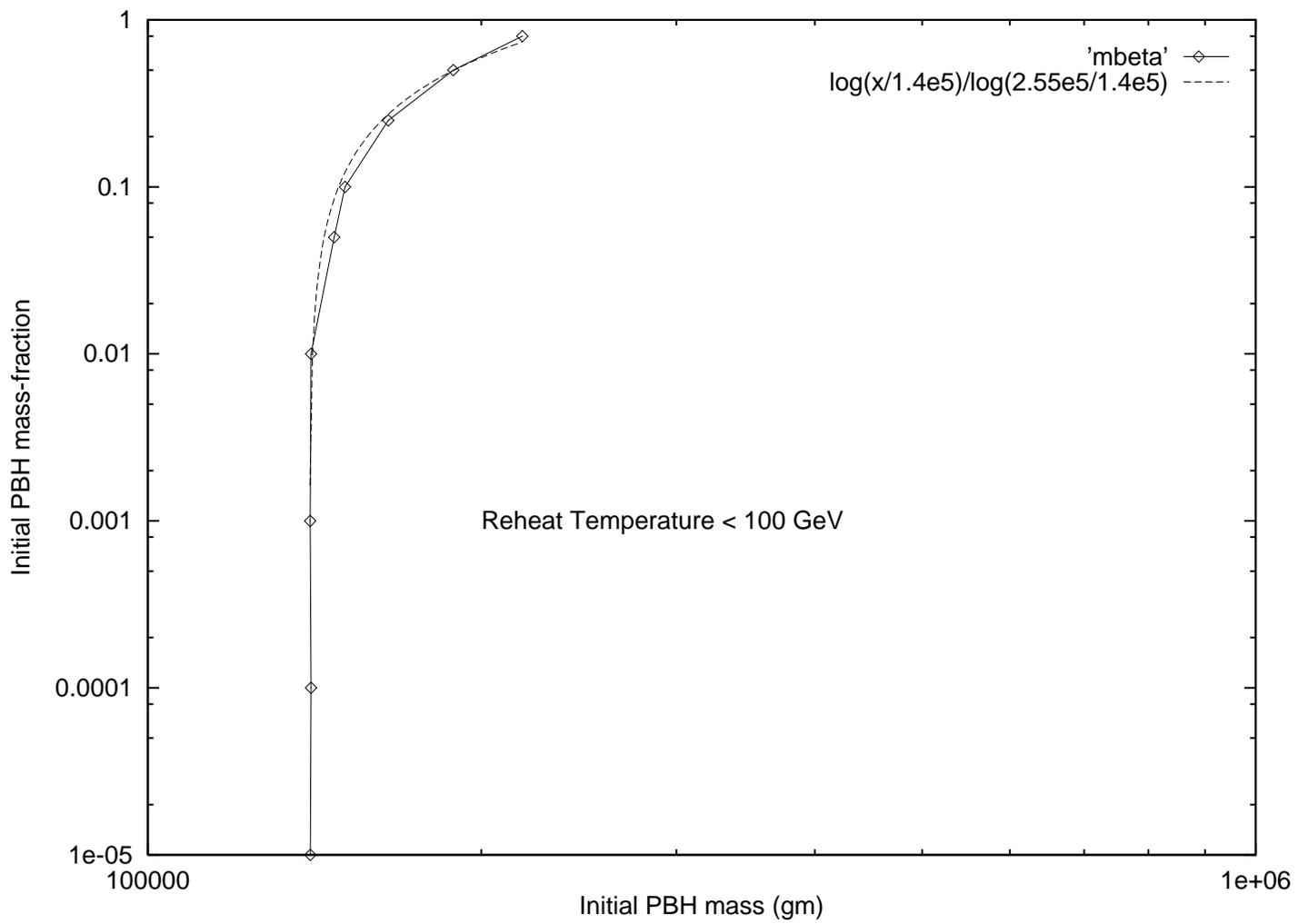}
   \caption {The combinations $\zeta$ and $m_0$ for which the reheat
      temperature $ = T_{EWPT} = 100$ GeV.
      The region with {\em acceptable} reheat temperatures $ <
      100$ GeV is indicated in the figure.
      The analytical fit with dotted line is {\em purely} empirical.}
   \label {fig:mbeta}
\end {figure}
\newpage
\begin {figure}[ht]
 \vskip 15truecm
      \includegraphics{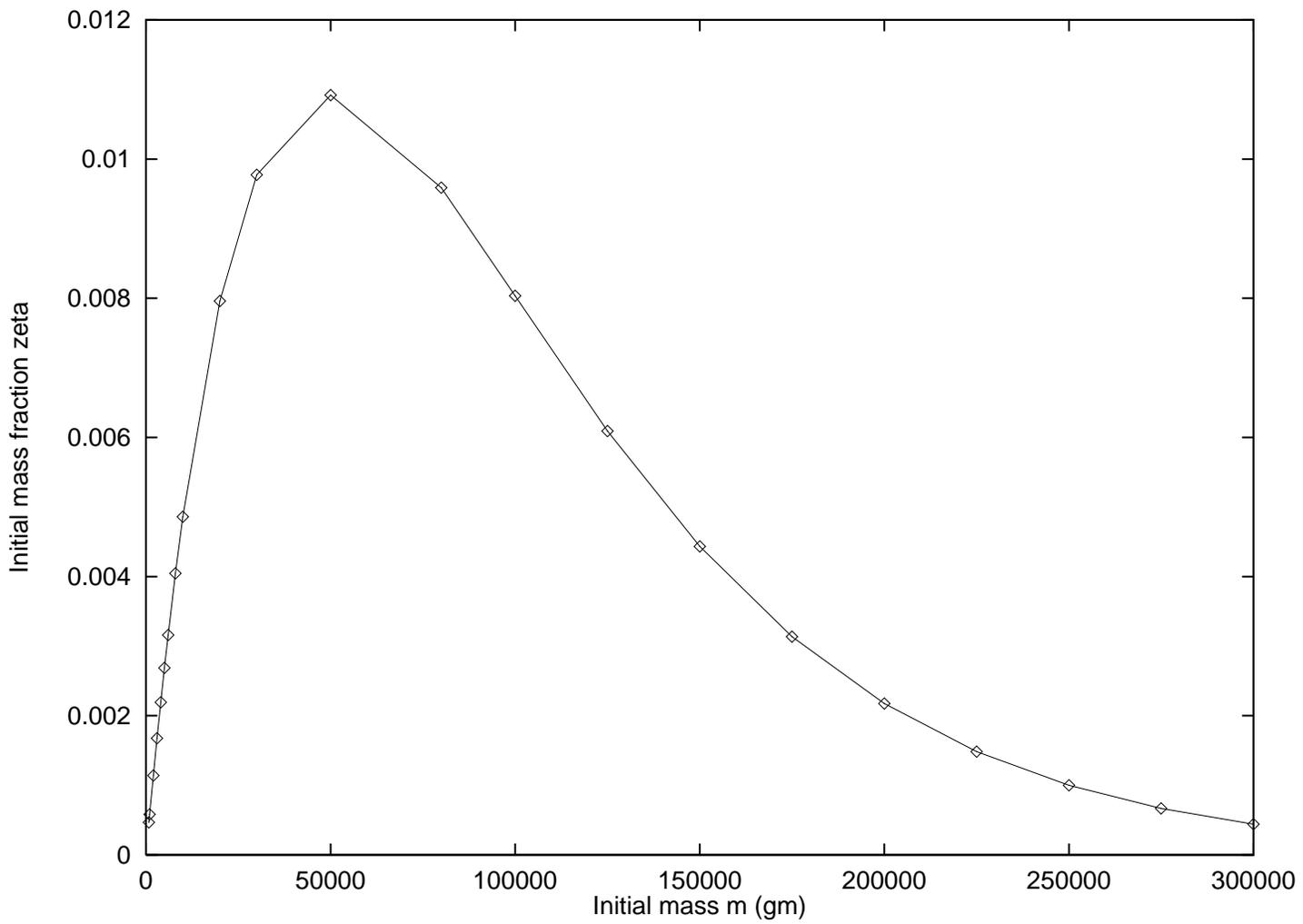}
   \caption {Black-hole mass spectrum: plot of $\zeta_i$ against
             $m_i(t_0)$.
             }
   \label {fig:mzeta}
\end {figure}
\newpage
\begin {figure}[ht]
 \vskip 15truecm
      \includegraphics{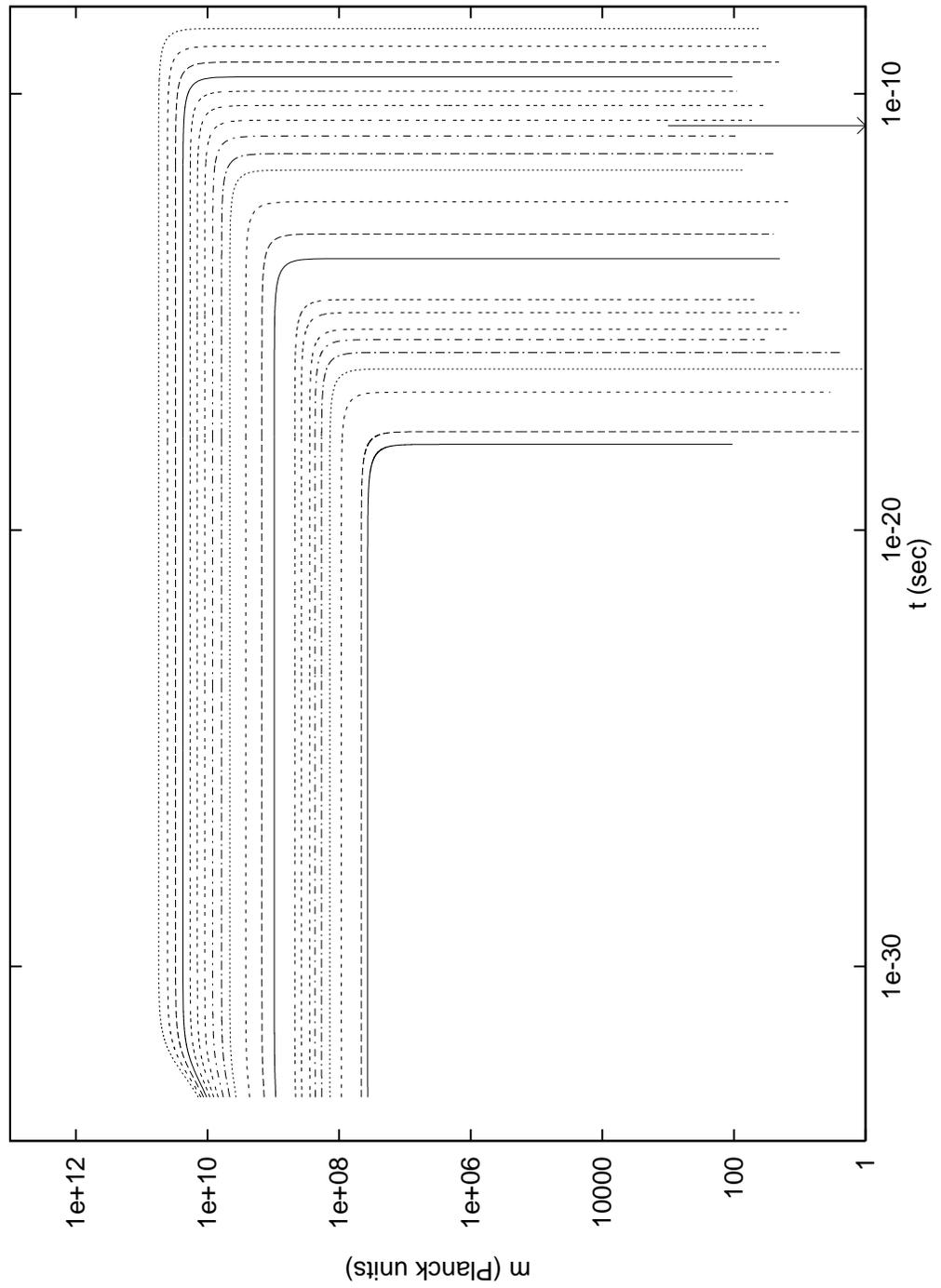}
   \caption {The evolution of the masses $m_i(t)$ of a collection
            of PBH masses distributed according to the spectrum shown
            in  fig \myref{fig:mzeta}.
            }
   \label {fig:spectrum}
\end {figure}
\newpage
\begin {figure}[ht]
 \vskip 15truecm
      \includegraphics{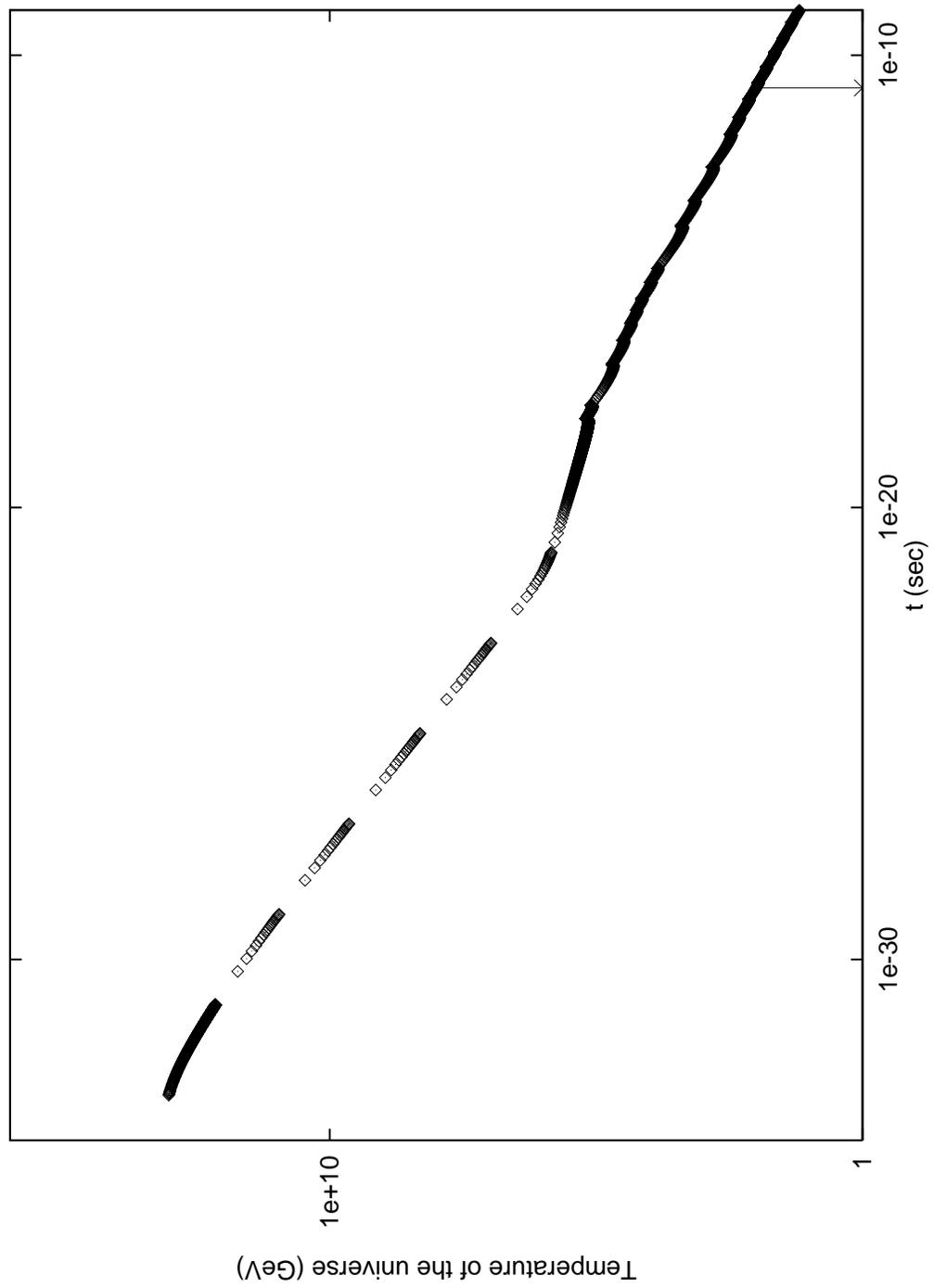}
   \caption {The cooling of the universe for the case where PBH
             masses are distributed according to the spectrum
             displayed in fig \myref{fig:mzeta}.
             The epoch of EWPT is marked by an arrow, and it
             takes place at $1.9\times 10^{-11}$ sec.
            }

   \label {fig:Tuni}
\end {figure}
\end{document}